\RequirePackage{etoolbox}

\newtoggle{preprint}
\toggletrue{preprint}
\newtoggle{review}
\togglefalse{review} %

\iftoggle{review}{
    \documentclass[nonacm,acmtog,anonymous]{acmart}
}{
    \documentclass[nonacm,acmtog]{acmart}
}

\usepackage{import}
\usepackage[ruled]{algorithm2e} %

\SetAlFnt{\small}
\SetAlCapFnt{\small}
\SetAlCapNameFnt{\small}
\SetAlCapHSkip{0pt}
\SetKw{OR}{or}
\SetKw{AND}{and}

\usepackage{xcolor,amsmath,tikz}
\usetikzlibrary{shapes}
\usepackage{amsmath}
\nottoggle{preprint}{
    \usepackage{amssymb}
}{}
\usepackage{bm}
\usepackage{ upgreek }
\usepackage{listings}
\usepackage[normalem]{ulem}
\usepackage{wrapfig}
\usepackage{xspace}
\usepackage{subcaption}
\usepackage{shellesc}
\usepackage{stackrel}
\usepackage{tcolorbox}
\usepackage{hyperref}
\usepackage{graphicx} %

\usepackage{pgf}
\usepackage{tikz}
\usetikzlibrary{
  calc,
  decorations.pathreplacing,
  decorations.pathmorphing,
  calligraphy,
  calc
}
\tikzstyle{arrow} = [very thick,->,>=stealth]

\newcommand{\mathdefault}[1][]{}
\usepackage[outline]{contour}
\usepackage{trimclip}

\usepackage{booktabs}
\usepackage{array}
\usepackage{multirow}

\usepackage[capitalise]{cleveref}
\if@cref@capitalise
\crefname{pluralneq}{Ineqs.}{Ineqs.}
\crefname{pluraleq}{Eqs.}{Eqs.}
\crefname{neq}{Ineq.}{Ineq.}
\else
\crefname{pluralneq}{ineqs.}{ineqs.}
\crefname{pluraleq}{eqs.}{eqs.}
\crefname{neq}{ineq.}{ineq.}
\fi
\makeatother
\Crefname{pluralneq}{Ineqs.}{Ineqs.}
\Crefname{pluraleq}{Eqs.}{Eqs.}
\Crefname{neq}{Ineq.}{Ineq.}
\creflabelformat{pluralneq}{#2(\textup{#1})#3}
\creflabelformat{pluraleq}{#2(\textup{#1})#3}
\creflabelformat{neq}{#2(\textup{#1})#3}

\iftoggle{preprint}{
    
}{
       
}

\usepackage{enumerate}
\usepackage{siunitx}

\graphicspath{{images/}}

\newcommand{\mv}[1]{\mathbf{#1}}

\newcommand{\dfdx}[2]{\ensuremath \partial_{#2} #1}
\newcommand{\ambient}{\mathbb{R}^3}

\newcommand{\Sp}{S}
\newcommand{\p}{\mv{p}}

\newcommand{\pn}{\mv{p}'}
\newcommand{\getpn}{\mv{r}}

\iftoggle{preprint}{
    \newcommand{\solid}{\boldsymbol{\omega}}
    \newcommand{\solidn}{\boldsymbol{\omega}'}
} {
    \newcommand{\solid}{\omega}
    \newcommand{\solidn}{\omega'}
}
\newcommand{\pos}{\mathbf{x}}

\newcommand{\posS}{\mathbf{p}}  %
\newcommand{\hitS}{\mathbf{r}} %
\newcommand{\sphere}{{\mathcal{S}^{2}}}

\newcommand{\mitsuba}{\textsc{Mitsuba~3}\xspace}
\newcommand{\redner}{\textsc{redner}\xspace}
\newcommand{\psdr}{\textsc{psdr-jit}\xspace}

\definecolor{codeback}{rgb}{0.97,0.97,0.97}
\definecolor{BrickRed}{HTML}{B6321C}
\lstdefinestyle{mypython}{
  language=Python,
  basicstyle=\footnotesize\ttfamily,
  keepspaces=true,
  numbers=left,
  numbersep=8pt,
  tabsize=4,
  frame=single,
  captionpos=b,
  numberstyle=\scriptsize\color{gray},
  backgroundcolor=\color{codeback},
  stringstyle=\color{Purple},
  commentstyle=\color{ForestGreen},
  keywordstyle=\color{BrickRed}\bfseries,
  mathescape=true,
  morekeywords={mitsuba,drjit},
}

\usepackage{tcolorbox}
\usepackage[frozencache,cachedir=minted-cache]{minted}
\usepackage{mdframed}
\usepackage{setspace}
\newminted{python}{autogobble,mathescape=true,escapeinside=||,fontsize=\small,linenos}

\BeforeBeginEnvironment{minted}{
    \begin{tcolorbox}[arc=1pt,colback=black!3!white,colframe=black!16!white,boxrule=0.3mm,before skip=1mm,after skip=1mm,top=.5mm,bottom=.5mm,left=1mm,right=1mm]
    }
\AfterEndEnvironment{minted}{\end{tcolorbox}}

\makeatletter
\newcommand{\printfnsymbol}[1]{%
  \textsuperscript{\@fnsymbol{#1}}%
}
\makeatother

\title{Radiative Backpropagation with Non-Static Geometry}

\nottoggle{review}{
    \author{Markus Worchel}
    \orcid{0000-0002-3469-6750}
    \authornote{Equal contribution.}
    \affiliation{%
       \institution{Technische Universität Berlin (TU Berlin)}
       \city{Berlin}
       \country{Germany}}
    \email{m.worchel@tu-berlin.de}
    
    \author{Ugo Finnendahl}
    \orcid{0000-0002-7098-1524}
    \authornotemark[1]
    \affiliation{%
       \institution{Technische Universität Berlin (TU Berlin)}
       \city{Berlin}
       \country{Germany}}
    \email{finnendahl@tu-berlin.de}

    \author{Marc Alexa}
    \orcid{0000-0002-9854-8466}
    \affiliation{%
       \institution{Technische Universität Berlin (TU Berlin)}
       \city{Berlin}
       \country{Germany}}
    \email{marc.alexa@tu-berlin.de}
}{}

\begin{document}

\begin{abstract}

Radiative backpropagation-based (RB) methods efficiently compute reverse-mode derivatives in physically-based differentiable rendering by simulating the propagation of differential radiance. A key assumption is that differential radiance is transported like normal radiance. We observe that this holds only when scene geometry is static and demonstrate that current implementations of radiative backpropagation produce biased gradients when scene parameters change geometry. In this work, we derive the differential transport equation without assuming static geometry. An immediate consequence is that the parameterization matters when the sampling process is not differentiated: only surface integrals allow a local formulation of the derivatives, i.e., one in which moving surfaces do not affect the entire path geometry. While considerable effort has been devoted to handling discontinuities resulting from moving geometry, we show that a biased interior derivative compromises even the simplest inverse rendering tasks, regardless of discontinuities.
An implementation based on our derivation leads to systematic convergence to the reference solution in the same setting and provides unbiased RB interior derivatives for path-space differentiable rendering.

\nottoggle{preprint}{

}{}

\end{abstract}


\maketitle

\section{Introduction}
\label{sec:intro}

In physically-based differentiable rendering, one is interested in differentiating the rendering equation~\cite{Kajiya1986}
\begin{equation}
    L_o(\posS, \solid) = L_e(\posS, \solid) +
    \int_{\sphere} L_i(\posS, \solid_i)\, f(\posS, \solid, \solid_i) \, \text{d} \solid_i^\bot, \label{eq:scattering}
\end{equation}
where $L_e$ is the emitted radiance from a surface point $\posS$ into a direction $\solid$, $f$ is the bidirectional scattering distribution function (BSDF), and the integral is over the sphere $\sphere$, with $\solid_i^\bot$ denoting a projected solid angle. The outgoing radiance $L_o$ and the incoming radiance $L_i$ are related via the transport relation
\begin{equation}
    L_i(\posS, \solid) = L_o(\hitS(\posS, \solid), -\solid), \label{eq:transport}
\end{equation}
and $\hitS(\posS, \solid)$ is the nearest surface point from $\posS$ in direction $\solid$.

The derivative is typically computed w.r.t.\ an arbitrary scene parameter $\pi$. If $\pi$ alters the scene geometry, its variation is often discussed in the context of discontinuity handling, as changes in geometry may alter occlusion relations. While $\pi$ is explicitly allowed to affect geometry, discontinuity handling is orthogonal to the main aim of this work. A scene may well have occlusion boundaries that do not affect the integrand, or that only marginally contribute to the derivatives when the geometry moves. For now, we assume such scenes and defer a discussion of discontinuities. 

Gradient-based optimization of scene parameters commonly relies on reverse-mode derivatives. Nimier-David et al.~\shortcite{NimierDavid2020Radiative} and Stam~\shortcite{Stam:2020:Adjoint} observe that these can be computed using the adjoint method, leading to \emph{radiative backpropagation} (RB): instead of building a complex computation graph for automatic differentiation (AD), derivatives are explicitly computed by simulating the transport of \emph{differential radiance}. First, differentiating both sides of the rendering equation \eqref{eq:scattering} yields \emph{differential scattering}:
\begin{multline}
    \dfdx{L_o}{\pi}(\posS, \solid) = \underbrace{\dfdx{L_e}{\pi}(\posS, \solid)}_{\text{Term 1}}\\
    + \int_{\sphere}  \big[ \underbrace{\dfdx{L_i}{\pi}(\posS, \solid_i) f(\posS, \solid, \solid_i)}_{\text{Term 2}} \\
      + \underbrace{L_i(\posS, \solid_i) \dfdx{f}{\pi}(\posS, \solid, \solid_i)}_{\text{Term 3}} \big] \, \text{d} \solid_i^\bot. \label{eq:diff_scattering_nimier}
\end{multline}
The three terms are interpreted as follows:
\begin{itemize}
    \item \textbf{Term 1}: Differential radiance is emitted by light sources if $\pi$ affects the emission (e.g. if $\pi$ is the brightness).
    \item \textbf{Term 2}: Differential radiance scatters like normal radiance according to the BSDF $f$
    \item \textbf{Term 3}: Differential radiance is emitted when the BSDF changes with $\pi$ (e.g. if $\pi$ affects the albedo of a surface).
\end{itemize}
Second, differentiating the transport equation \eqref{eq:transport} yields \emph{differential transport}, presented by Nimier-David et al.~\shortcite{NimierDavid2020Radiative} as
\begin{equation}
\label{eq:diff_transport_nimier}
    \dfdx{L_i}{\pi}(\posS, \solid) = \dfdx{L_o}{\pi}(\hitS(\posS, \solid), -\solid),
\end{equation}
i.e., differential radiance is transported like normal radiance. 

This differential transport relation, however, holds only for the case of \emph{static} geometry, i.e., the scene geometry is unaffected by $\pi$, yet this assumption and the consequences are not commonly discussed. In fact, we derive the differential transport equation for \emph{non-static} geometry and show, with simple experiments, that existing RB implementations consider only a subset of the arising derivatives (\cref{sec:observation}). This behavior can be traced back to the theory of RB with ``detached sampling'', which, however, is incomplete in the sense that it only considers sampling of \emph{directions} with \emph{static} geometry. We derive the theory for non-static geometry and show that RB with detached sampling leads to non-local derivatives, depending on how the rendering integral is parameterized (\cref{sec:relation}). We verify our theory and implementation on several examples and show that it leads to an RB approach for the interior derivatives in path-space differentiable rendering (\cref{sec:experiments}).

\section{Differential Transport and Geometry}
\label{sec:observation}

The differential transport equation \eqref{eq:diff_transport_nimier} hides an intricate relation: when geometry is non-static, the nearest surface point $\hitS(\posS, \solid, \pi)$ can depend on $\pi$. Notice that the points $\posS$ and $\hitS$ are to be understood as points in a (hypothetical) global \emph{parametrization} $\mathcal{M} \subset \mathbb{R}^2$ of the surface, which is embedded in the \emph{ambient space} $\ambient$. We denote the mapping from a point in the parameterization $\posS \in \mathcal{M}$ to the ambient space by $S(\posS, \pi) \in \ambient$, which naturally depends on the geometry.

A change of the intersection point $\hitS \in \mathcal{M}$ with $\pi$ can now arise in two ways: (1) the surface at $\posS$ evolves in the ambient space, such that the origin $S(\posS, \pi)$ of the ray cast to determine $\hitS$ moves; and (2) the surface at $\hitS$ evolves in ambient space.

The potential movement of the intersection point with $\pi$ must be considered when differentiating the transport equation \eqref{eq:diff_transport_nimier}. Applying the multivariate chain rule to the equation, with an explicit dependency on $\pi$ for completeness, yields:
\begin{multline}
    \dfdx{L_i}{\pi}(\posS, \solid, \pi) = \dfdx{L_o}{\pi}(\hitS(\posS, \solid, \pi_0), -\solid, \pi)\\
    + \dfdx{L_o}{\posS}(\hitS(\posS, \solid, \pi_0), -\solid, \pi_0) \dfdx{\hitS}{\pi}(\posS, \solid, \pi),\label{eq:diff_transport_chain_rule}
\end{multline}
where $\pi_0 = \pi$. Differential radiance is \emph{not just} transported like normal radiance (first term) but differential radiance is also generated through the transport (second term). %
For example, consider an area light source with an emission texture, and let $\pi$ be the position of the light. At each point on the light source, the emission is spatially varying because of the texture ($\dfdx{L_o}{\posS} \neq 0$), and the intersection point of a ray with the light source can depend on $\pi$ ($\dfdx{\hitS}{\pi} \neq 0$). Therefore, any path segment that terminates on the light could generate differential radiance. Importantly, transport between \emph{any} surfaces, including non-emissive ones, could generate differential radiance: 

\begin{quote}
\emph{Differential radiance is emitted through transport from and to geometry that moves with $\pi$}.
\end{quote}

To see why, we differentiate the scattering equation \eqref{eq:scattering} w.r.t.\ the position $\posS$
\begin{multline}
    \dfdx{L_o}{\posS}(\posS, \solid) = \underbrace{\dfdx{L_e}{\posS}(\posS, \solid)}_{\text{Term A}}\\
    + \int_\sphere \big[ \underbrace{\dfdx{L_i}{\posS}(\posS, \omega_i)\, f(\posS, \solid, \solid_i)}_{\text{Term B}}\\
     +\underbrace{L_i(\posS, \omega_i)\, \dfdx{f}{\posS}(\posS, \solid, \solid_i)}_{\text{Term C}} \big]  \, \solid_i^\bot. \label{eq:diff_scattering_position}
\end{multline}
Given that the intersection point is affected by $\pi$, i.e., $\dfdx{\hitS}{\pi} \neq \mv{0}$, the three terms in this equation have the following interpretation regarding the product $\dfdx{L_o}{\posS} \dfdx{\hitS}{\pi}$ from Equation~\eqref{eq:diff_transport_chain_rule}
\begin{itemize}
    \item \textbf{Term A}: Differential radiance is emitted by the transport from light sources if their emission is spatially varying (e.g. an area light with an emission texture).
    \item \textbf{Term B}: Differential radiance is emitted by the transport from surfaces if the \emph{incoming radiance} $L_i$ is spatially varying.
    \item \textbf{Term C}: Differential radiance is emitted by the transport from surfaces if their BSDF is spatially varying (e.g. a surface with an albedo texture).
\end{itemize}
Term B has a recursive nature, revealed by differentiating the transport equation, where $\posS_0$ has the value of $\posS$:
\begin{equation}
    \dfdx{L_i}{\posS}(\posS, \solid) = \dfdx{L_o}{\posS}(\hitS(\posS_0, \solid), -\solid) \dfdx{\hitS(\posS, \solid)}{\posS}. \label{eq:diff_transport_position}
\end{equation}
In practice, the incident radiance at \emph{every} surface point will almost surely be spatially varying as radiance is reflected towards a point from various other surfaces: as soon as scene geometry moves with $\pi$, differential radiance is generated through transport, regardless of whether there are spatially varying BSDFs or emission.

\subsection{Case Study} %
\label{sec:case_1}

\begin{figure}
    \centering
    \includegraphics[width=\linewidth, trim=0 0 20 0]{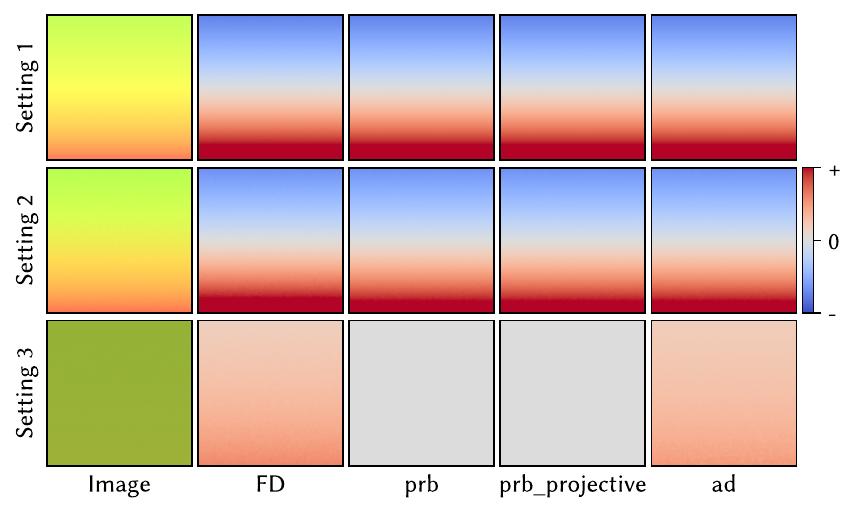}
    \caption{Three scenes with a backward-moving rectangle -- spatially-varying emission (first row), spatially varying BSDF (second row), and indirect illumination (third row) -- are rendered (first column). The derivatives of the red channel are displayed: finite differences (FD) (second column), PRB (third column), and the PRB variant of projective sampling (fourth column). All PRB methods compute incorrect zero derivatives for the last setting. Automatic differentiation (ad) yields correct derivatives (last column).}
    \label{fig:case_1}
    \iftoggle{preprint}{
        \Description[]{} %
    }{}
\end{figure}

We observe that existing RB implementations consider only Terms A and C and omit Term B. Consequently, the geometry derivatives will be biased, even for surfaces directly observed by a camera.

\begin{wrapfigure}{r}[12pt]{0.34\linewidth}
    \nottoggle{preprint}{\vspace{-10pt}}{}
    \hspace{-15pt}
    \includegraphics[width=\linewidth]{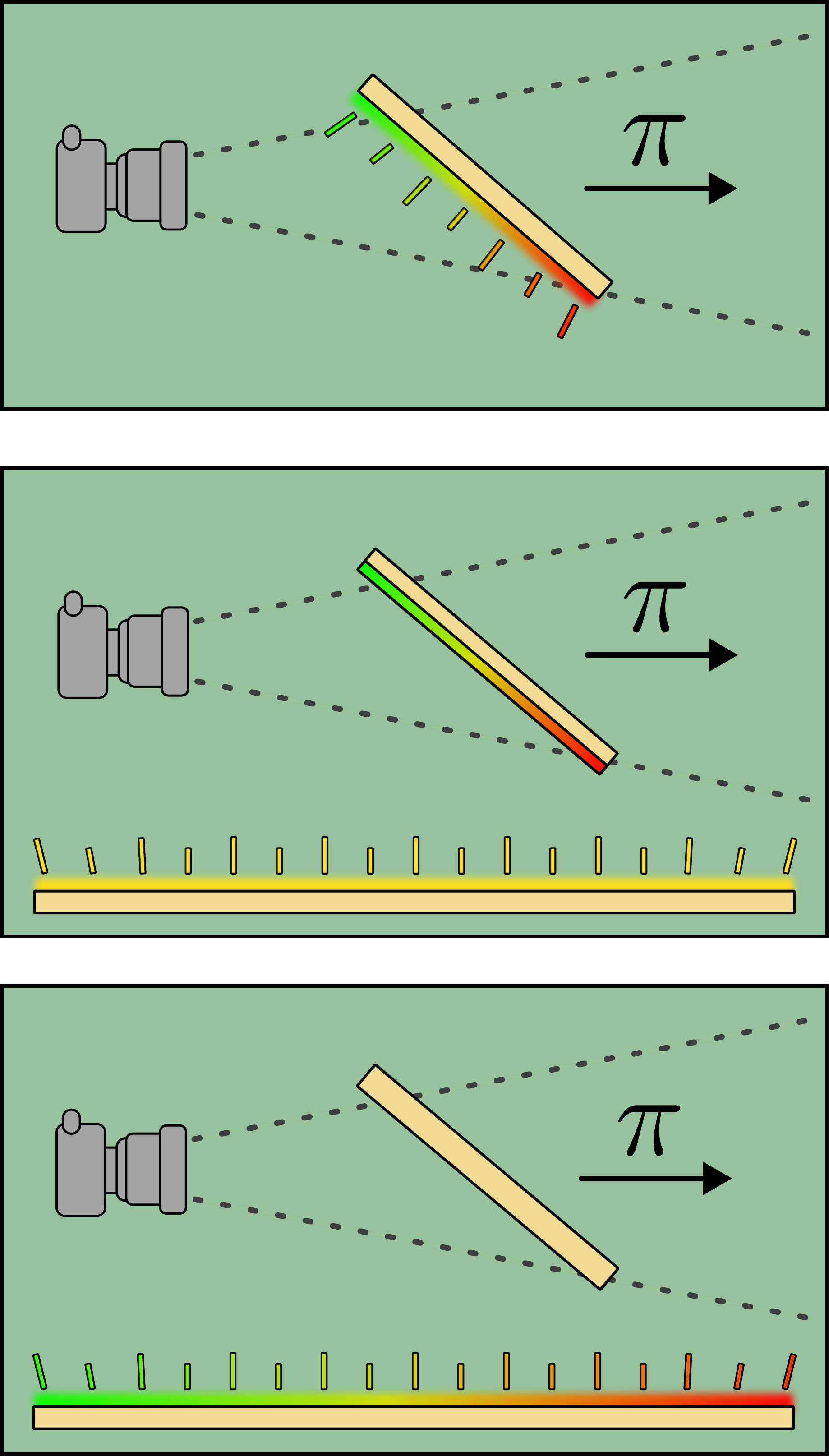}
    \vspace{-10pt}
\end{wrapfigure}
We demonstrate this behavior for the different RB-based integrators in \mitsuba~\cite{mitsuba} (version 3.6.4) using a set of simple examples. In the first setting (inset, top) the camera observes a slab with a spatially varying emission, which is moved away from the camera by $\pi$. In the second setting (inset, middle) the camera observes the same slab, which now has a spatially varying diffuse albedo and is illuminated from below. In the third setting (inset, bottom), the moving slab has a constant albedo but is illuminated from below by another slab with spatially varying emission. Notice that, even though geometry is allowed to move, the integrand is free of discontinuities in the first setting and discontinuities only marginally contribute to the derivative in the last two settings (the bottom slab is larger than shown).

For the derivatives w.r.t.\ $\pi$, one must account for Term A in the first setting and Term C in the second setting. Indeed, \mitsuba's RB integrators compute these derivatives (Figure~\ref{fig:case_1}, rows 1 and 2). In the third setting, Term B has a noticeable contribution: the observed slab is illuminated by a spatially varying light, so the incident radiance changes when moving from one point on the surface to another ($\dfdx{L_i}{\posS} \neq 0$ on the first slab). The RB integrators do \emph{not} account for this (Figure~\ref{fig:case_1}, last row). This even affects the integrators designed for geometry optimization, such as \verb|prb_projective|, which implements RB-style projective sampling~\cite{Zhang:2023:Projective}.

On a technical level, the reason why the implementation considers Terms A and C, even though the existing theory misses the transport-related differential radiance, is that the intersection point 
is attached, which means gradient tracking with respect to the intersected surface is enabled. Since the emission at the intersection and the BRDF are evaluated with an attached intersection point, the subsequent reverse accumulation automatically includes $\dfdx{L_e}{\posS} \dfdx{\hitS}{\pi}$ (Term A) and $L_i \dfdx{f}{\posS} \dfdx{\hitS}{\pi}$ (Term C). However, the ray defining the next path segment is detached from the intersection point, so changes in incident radiance are not included (Term B).

\paragraph*{PRB and automatic differentiation}

The result of path replay backpropagation (PRB)~\cite{Vicini2021} (\verb|prb|), the most common variant of RB, should match automatic differentiation (\verb|ad|, with detached sampling (\cref{sec:detached_directions})) by design, but it does not in the last setting (\cref{fig:case_1}). The reason is that automatic differentiation includes Term B, which closely matches finite differences. This is not an issue of replaying paths but of the underlying theory for static geometry.

\section{Path Replay Backpropagation for Non-Static Geometry}
\label{sec:relation}

The most common variant of radiative backpropagation is path replay backpropagation (PRB)~\cite{Vicini2021}, which ``replays'' the light paths from the forward simulation when simulating the transport of differential radiance. Typically, (P)RB is considered with \emph{detached sampling}, i.e., the sampling process and probability densities in Monte Carlo estimators are not differentiated. Vicini et al.~\shortcite{Vicini2021} show that this leads to a local method that does not need to consider perturbations of path geometry with the scene parameters. %
However, the resulting derivatives are only local if the geometry is \emph{static} (\cref{sec:detached_directions}).

As we show in \cref{sec:detached_positions}, a truly local approach for derivatives with non-static geometry requires a theory for detached sampling of positions in the \emph{three-point form}. Existing literature only provides an RB theory for (detached) sampling of directions and only for static geometry. We believe that this theoretical gap has led to current implementations mixing the different perspectives, without consistently following either one. 

In conclusion, we make two central observations that, to the best of our knowledge, are absent in the current literature: 

\begin{enumerate}
    \item \cref{sec:detached_directions}: With non-static geometry, global path geometry must be considered for direction sampling \emph{even with detached sampling}; Vicini et al.~\shortcite{Vicini2021} state that this is a unique to \emph{attached} sampling, but this is only true if geometry is static.
    \item \cref{sec:detached_positions}: When taking the perspective of the three-point form, detached sampling makes the PRB approach local but one must correctly account for the (derivatives of the) transformation determinants and the intersection point.
\end{enumerate}

\subsection{Detached Sampling of Directions}
\label{sec:detached_directions}

Vicini et al.~\shortcite{Vicini2021} introduce the notion of ``attached'' and ``detached'' sampling for importance sampling of an integral
\begin{equation}\label{eq:spherical}
    \int\limits_\sphere L(\posS, \solid, \pi) \, \text{d} \solid,
\end{equation}
which represents the scattering equation \eqref{eq:scattering} with a spherical integration domain, i.e., $L$ contains the emission $L_e$, the BSDF $f$, the incoming radiance $L_i$ and the foreshortening cosine term. Importance sampling is cast as uniform sampling from the unit hypercube $\mathcal{U}$ with a change of variables
\begin{equation}\label{eq:attached}
    \int\limits_{\mathcal{U}} \frac{L\big(\posS, T(\mv{u}, \pi), \pi \big)}{p\big(T(\mv{u}, \pi), \pi\big)} \, \text{d} \mv{u},
\end{equation}
where $p(\solid, \pi)$ is the target density from which the mapping $T(\mv{u}, \pi) = \solid$ is constructed~\cite{Vicini2021}. Both typically depend on the scene parameter $\pi$ (e.g. on the roughness of a material).

Differentiating the integral yields an \emph{attached} differential sampling strategy
\begin{equation}
    \int\limits_{\mathcal{U}} \dfdx{\left[ \frac{L\big(\posS, T(\mv{u}, \pi), \pi \big)}{p\big(T(\mv{u}, \pi), \pi\big)} \right]}{\pi} \, \text{d} \mv{u}
\end{equation}
that tracks how changes in $\pi$ affect the sampling process. Let $\pi_0$ be a parameter that matches the value of $\pi$ but does not participate in differentiation. A \emph{detached} differential sampling strategy is
\begin{equation}
    \int\limits_{\mathcal{U}} \frac{\dfdx{L\big(\posS, T(\mv{u}, \pi_0), \pi \big)}{\pi}}{p\big(T(\mv{u}, \pi_0), \pi_0\big)} \, \text{d} \mv{u}.
\end{equation}
Important and maybe surprising: both attached and detached strategies are valid estimators of the derivative as long as all points with $\dfdx{L}{\pi} \neq 0$ are sampled with nonzero probability and the integrand is free of Dirac deltas~\cite{Zeltner2021}.

\paragraph*{Non-static geometry moves a path}

\begin{wrapfigure}{r}[12pt]{0.3\linewidth}
    \vspace{-10pt}
    \hspace{-15pt}
    \includegraphics[width=\linewidth]{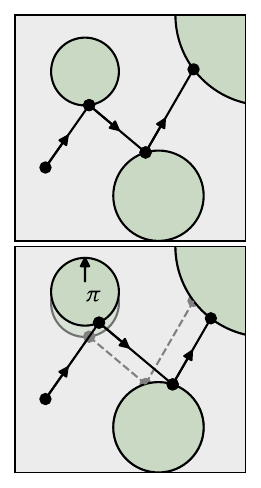}
    \vspace{-10pt}
\end{wrapfigure}
The samples generated by the mapping $T$ are \emph{directions} and therefore, in detached sampling, $\solid_0 = T(\mv{u}, \pi_0)$ is a \emph{direction} independent of $\pi$. If directions along a path are sampled independent of any parameter, then the movement of intersected geometry with (another) parameter $\pi$ can lead to the movement of successive path vertices: assume $\posS$ is an intersection point on a surface that moves with $\pi$. Since the sampled direction $\solid_0$ is constant, the next intersection point $\hitS(\posS, \solid_0)$ of a ray originating from $\posS$ into direction $\solid_0$ will depend on $\pi$, because $\pi$ affects the ray origin in ambient space $S(\posS, \pi)$. As the next intersection point depends on $\pi$, the ones after will as well by recursion (see inset).

\paragraph*{Computing the non-local term}

The connection between the derivative and global path geometry is captured by Term B in Eq.~\eqref{eq:diff_scattering_position} and its recursion in Eq.~\eqref{eq:diff_transport_position}, which, as shown in \cref{sec:observation}, existing path replay implementations do not account for. Strictly following the theory of detached direction sampling, one can recover the missing contribution. This leads to a construction very similar to attached sampling, where global path geometry is relevant, with the main difference being that the probability density and the sampled direction are not differentiated (see Appendix~\ref{sec:attached_as_detached}).

\subsection{Detached Sampling of Positions}
\label{sec:detached_positions}

The three-point form of the scattering equation~\cite{Veach1998} emerges when the integral over the spherical domain (\cref{eq:spherical}) is reparameterized as an integral over surface points in the scene
\begin{equation}\label{eq:three_point_form}
    \int\limits_\mathcal{M} L\left(\posS, S(\posS, \pi) \to S(\posS', \pi), \pi \right)\, D(\posS, \posS', \pi) \, \text{d} \posS'.
\end{equation}
$D$ represents the Jacobian determinant of the reparameterization
\begin{equation}
    D(\posS, \posS', \pi) = G\big(\posS, \posS', \pi\big) \, \vert \dfdx{S}{\posS_u}(\posS', \pi) \times \dfdx{S}{\posS_v}(\posS', \pi) \vert,
\end{equation}
with the geometry term
\begin{equation}
    G(\posS, \posS', \pi) = V(\posS, \posS', \pi)\, \frac{\vert \mv{n}(\posS', \pi)^\top \left(S(\posS', \pi) \to S(\posS, \pi)\right) \vert}{\Vert S(\posS, \pi) - S(\posS', \pi) \Vert^2},
\end{equation}
where $V$ is the binary visibility between the two points, $\mv{n}$ the surface normal at $\posS'$ and $\mv{x} \to \mv{y}$ the direction from $\mv{x} \in \ambient$ to $\mv{y} \in \ambient$. Notice that this is the intrinsic form, over a global parameterization $\mathcal{M} \subset \mathbb{R}^2$, as discussed in \cref{sec:observation}.

A detached sampling strategy for positions can be derived as in Section~\ref{sec:detached_directions}, by reparameterizing the integral over the unit cube 
\begin{equation}
    \int\limits_\mathcal{U} \frac{L\big(\posS, S(\posS, \pi) \to S(T(\mv{u}, \pi), \pi), \pi\big) \, D\big(\posS, T(\mv{u}, \pi), \pi\big)}{p\big(T(\mv{u}, \pi), \pi\big)} \, \text{d} \mv{u},
\end{equation}
differentiating and substituting $\pi_0$ that does not participate in differentiation
\begin{equation}
    \int\limits_\mathcal{U} \frac{\dfdx{\big[ L\big(\posS, S(\posS, \pi) \to S(T(\mv{u}, \pi_0), \pi), \pi\big)\, D\big(\posS, T(\mv{u}, \pi_0), \pi\big) \big]}{\pi}}{p\big(T(\mv{u}, \pi_0), \pi_0 \big)} \, \text{d} \mv{u}. \label{eq:detached_position_sampling}
\end{equation}

\paragraph*{Breaking the recursion of differential transport}
\label{sec:breakingtherecursion}

To see why this reparameterization makes the derivatives local, consider
\begin{wrapfigure}{r}[12pt]{0.49\linewidth}
    \vspace{-10pt}
    \hspace{-15pt}
    \includegraphics[width=\linewidth]{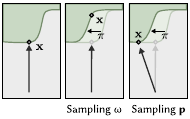}
    \vspace{-10pt}
\end{wrapfigure}
an ambient-space intersection point $\mv{x} = S(\hitS(\posS, \solid, \pi), \pi)$ that may be computed as part of the (recursive) function $L$. Sampling a direction $\solid_0$ as in Section~\ref{sec:detached_directions}, yields an intersection point $\mv{x} = S(\hitS(\posS, \solid_0, \pi), \pi)$ that slides along the surface when the intersected object or the ray origin moves (inset middle). On the other hand, sampling a static point $\posS_0 = T(\mv{u}, \pi_0)$ in the parameterization $\mathcal{M}$, yields an intersection point in ambient space
\begin{equation}
    \mv{x} = S \big(\hitS(\posS, S(\posS, \pi) \to S(\posS_0, \pi), \pi), \pi \big) = S(\posS_0, \pi)
\end{equation}
that moves \emph{with} the shape (inset right). Since the parameter position of the intersection point is independent of $\pi$, the differential transport equation \eqref{eq:diff_transport_chain_rule} becomes
\begin{multline}
\label{eq:diff_transport_threepoint}
    \dfdx{L_i}{\pi}\big(\posS, S(\posS, \pi) \to S(\posS_0, \pi), \pi\big) = \\\dfdx{L_o}{\pi}\big(\posS_0, S(\posS_0, \pi_0) \to S(\posS, \pi_0), \pi\big)\\
    + \dfdx{L_o}{\solid}\big(\posS_0, S(\posS_0, \pi_0) \to S(\posS, \pi_0), \pi_0\big) \dfdx{[S(\posS_0, \pi) \to S(\posS, \pi)]}{\pi}.
\end{multline}%
The differential $\dfdx{L_o}{\solid}$ is local because only the emission and the BRDF value at the intersection point $\posS_0$ vary with the direction $\solid$, so, differential transport requires only local path geometry.

\paragraph*{Contribution of individual samples}
\label{par:contribution_of_samples}

It may seem counterintuitive but the derivatives account for the spatially varying albedo and emission in the settings from \cref{sec:case_1}, despite samples being \emph{fixed} locations on the surface. Take, for example, Setting 2, where the camera observes an object with spatially varying albedo moving with $\pi$: when sampling directions, the derivative of a single sample w.r.t.\ $\pi$ captures the changing albedo, e.g., from red to green, through the product of $L_i \dfdx{f}{\posS}$ (Term C in $\dfdx{L_o}{\posS}$, \cref{eq:diff_scattering_position}) and $\dfdx{\hitS}{\pi}$ (\cref{eq:diff_transport_chain_rule}). When sampling positions, the albedo at a single sample is constant, but its derivative w.r.t.\ $\pi$ considers the \emph{weighting} of it through the product of $L_i \dfdx{f}{\solid}$ (part of $\dfdx{L_o}{\solid}$ from \cref{eq:diff_transport_threepoint}) and $\dfdx{[S(\posS_0, \pi) \to S(\posS, \pi)]}{\pi}$, as well as through the derivative of the Jacobian determinant $\dfdx{D}{\pi}$. 
The combination of all sample derivatives then yields a valid estimate of the differential integral. %

Notice that this also leads to different noise patterns and variance, making equal sample counts not necessarily comparable.

\paragraph*{Importance sampling and Jacobian determinant}

The derivative of the Jacobian determinant is easily missed when constructing a path using BSDF importance sampling. %
This is because the PDF for positions $p$ is typically derived from a PDF for directions $p_{\solid}$
\begin{equation}
p(\dots) = p_{\solid}(\dots) D(\dots),
\end{equation}
by warping it with the determinant $D$. It may seem that the determinants cancel when plugging the PDF into \cref{eq:detached_position_sampling} but this is not the case: the determinant in the numerator depends on $\pi$, while the determinant in the denominator is part of the PDF and therefore independent of $\pi$:
\begin{equation}\label{eq:d_over_d}
\frac{D\big(\posS, T(\mv{u}, \pi_0), \pi\big)}{D\big(\posS, T(\mv{u}, \pi_0), \pi_0\big)}.
\end{equation}
The fraction always evaluates to $1$, but its derivative w.r.t. $\pi$ is (generally) not $0$. This is also noted by Zhang et al.~\shortcite{Zhang:2020:PSDR}.

\paragraph*{Pseudocode}
\label{sec:pseudocode_threepoint}

\begin{listing}[t]
\begin{minted}[escapeinside=||,mathescape=true,fontsize=\footnotesize,linenos]{python}
def sample_path(|$\mv{p}_{cam}$|, |$\solid$|):
   |$L$| = 0, |$\beta$| = 1
   |$\mv{p}_p$| = |$\mv{p}_{cam}$|
   |$\mv{p}_c$| = intersect(to_ambient(|$\mv{p}_{cam}$|), |$\solid$|)
   for i in range(1, N+1):
      |$\pos_p$| = to_ambient(|$\mv{p}_p$|)
      |$\pos_c$| = to_ambient(|$\mv{p}_c$|)
        
      |$L$| += |$\beta$| * |$L_e(\posS_c, \pos_c \to \pos_p)$|
        
      |$\solid^{\prime}$|, bsdf_value, bsdf_pdf = sample_bsdf(|$\dots$|)
      bsdf_value = eval_bsdf(|$\posS_c$|,|$\pos_c\to \pos_p$|,|$\solid^{\prime}$|)  
      |$\beta$| *= bsdf_value / bsdf_pdf
      |$\mv{p}_p$| = |$\mv{p}_c$|
      |$\mv{p}_c$| = intersect(|$\pos_c$|, |$\solid^{\prime}$|)
   return |$L$|
\end{minted}
\vspace{0.05cm}
\begin{minted}[escapeinside=||,mathescape=true,fontsize=\footnotesize,linenos]{python}
def sample_path_adjoint(|$\mv{p}_{cam}$|, |$\solid$|, |$L$|, |$\delta L$|):
   |$\beta$| = |$1$|
   |$\mv{p}_p$| = |$\mv{p}_{cam}$|
   |$\mv{p}_c$| = detach(intersect(to_ambient(|$\mv{p}_{cam}$|), |$\solid$|))
   for i in range(1, N+1):
      |$\pos_p$| = to_ambient(|$\mv{p}_p$|)
      |$\pos_c$| = to_ambient(|$\mv{p}_c$|)
        
      # Derivative of L_e
      Le = |$\beta$| * |$L_e(\posS_c, \pos_c \to \pos_p)$|
      |$\delta_\pi$| += backward_grad(Le, |$\delta L$|)
        
      # Derivative of Jacobian determinant
      |$D$| = reparam_det(|$\posS_p, \posS_c$|)
      |$\delta_\pi$| += backward_grad(|$D$|, |$\delta L$| * |$L$| / |$D$|)

      |$L$| -= Le

      # Derivative of BSDF
      |$\solid^{\prime}$|, bsdf_pdf = sample_bsdf(|$\dots$|)
      |$\mv{p}_n$| = detach(intersect(|$\pos_c$|, |$\solid^{\prime}$|))
      |$\pos_n$| = to_ambient(|$\mv{p}_n$|)
      bsdf_value = eval_bsdf(|$\posS_c$|,|$\pos_c\to \pos_p$|,|$\pos_c\to \pos_n$|)
      |$\delta_\pi$| += backward_grad(bsdf_value, 
                          |$\delta L$| * |$L$| / bsdf_value)
                          
      |$\beta$| *= bsdf_value / bsdf_pdf
      |$\mv{p}_p$| = |$\mv{p}_c$|
      |$\mv{p}_c$| = |$\mv{p}_n$|
   return |$\delta_\pi$|
\end{minted}
\vspace{0.25cm}
\caption{Pseudocode of detached PRB in three-point form. We omit emitter sampling, which is straightforward to add using multiple importance sampling. No derivatives are tracked across loop iterations. The used functions are defined in \cref{sec:pseudocode_threepoint} and a theoretical derivation of the code can be found in \cref{sec:appendix}.}
\label{lst:pseudocode}
\end{listing}

In Listing~\ref{lst:pseudocode} we provide pseudocode for detached path replay backpropagation in the three-point form, following the style of the listings by Vicini et al.~\shortcite{Vicini2021}. The function $\texttt{sample\_path}$ computes the radiance $L$ measured by a path that originates from the camera at position $\posS_\text{cam}$ into direction $\solid$ (this direction is derived from a sample on the sensor). 
The radiance $L$ and the adjoint radiance $\delta L$ (the derivative of a loss function w.r.t.\ $L$) are then passed to $\texttt{sample\_path\_adjoint}$ to retrieve the derivative of the loss w.r.t.\ $\pi$.
The functions are defined as follows:
\begin{itemize}
    \item \texttt{intersect} returns the intersection point of a ray with the scene as a point in the surface parametrization $\mathcal{M}$
    \item \texttt{to\_ambient} transforms a point from the parametrization to ambient space $\ambient$, considering the scene parameters
    \item \texttt{detach} disables derivative tracking for the given variable
    \item \texttt{reparam\_det} returns the value of the reparameterization determinant $D$
    \item \texttt{sample\_bsdf} samples a direction from the current BSDF and also returns the PDF value of the sample
    \item \texttt{eval\_bsdf}$(\posS, \solid_o, \solid_i)$ computes the BSDF value $f(\posS, \solid_o, \solid_i)$
    \item \texttt{backward\_grad}($f,\frac{\partial g}{\partial f}$) uses reverse-mode differentiation to evaluate $\frac{\partial f}{\pi} \frac{\partial g}{\partial f}$
\end{itemize}
The code has two notable differences from the original, which uses detached sampling of directions. 

First, each sample $T(\mv{u}_i,\pi_0)$ is a surface point $\mv{p}_i$ in the 2D parameter space $\mathcal{M}$, \emph{independent} of $\pi$. For BSDF importance sampling this means that one first samples a direction (\verb|sample_bsdf(...)|), then intersects the next surface but does not differentiate the operation (\verb|detach(intersect(...))|). The surface point is transformed back into the ambient space $\mathbb{R}^3$ using the mapping $S(\mv{p}, \pi)$ (\verb|to_ambient(...)|); this transformation is differentiated because it potentially depends on $\pi$ (the sample position in $\mathbb{R}^3$ follows if $\pi$ moves the corresponding shape).

Second, the Jacobian determinant $D$ from the reparameterization must be differentiated as it might depend on $\pi$. The BSDF importance sampling PDF contains the factor $\frac{1}{D}$, but it is detached, so the derivative of their product is generally non-zero (c.f. \cref{eq:d_over_d}). By the product rule, the derivative $\dfdx{D}{\pi}$ must be multiplied by $L$ and, for reverse-mode differentiation, also by the adjoint radiance $\delta L$. This leads to another gradient contribution. In \texttt{sample\_path} the product of $D$ and $\frac{1}{D}$ simply cancels, so it can be omitted.

\section{Implementation and Experiments}
\label{sec:experiments}

We have implemented detached sampling in the three-point form (\cref{sec:detached_positions}) in \mitsuba{}~\cite{mitsuba} as an automatic differentiation variant (\verb|ad_threepoint|) and a path replay variant (\verb|prb_threepoint|). We have also implemented detached sampling of directions (\cref{sec:detached_directions}) but only an automatic differentiation variant (\verb|ad|). All integrators use multiple importance sampling, combining BSDF and emitter samples. \nottoggle{preprint}{The code for all experiments can be found in the supplementary material}.

\begin{figure}
    \centering
    \includegraphics[width=\linewidth]{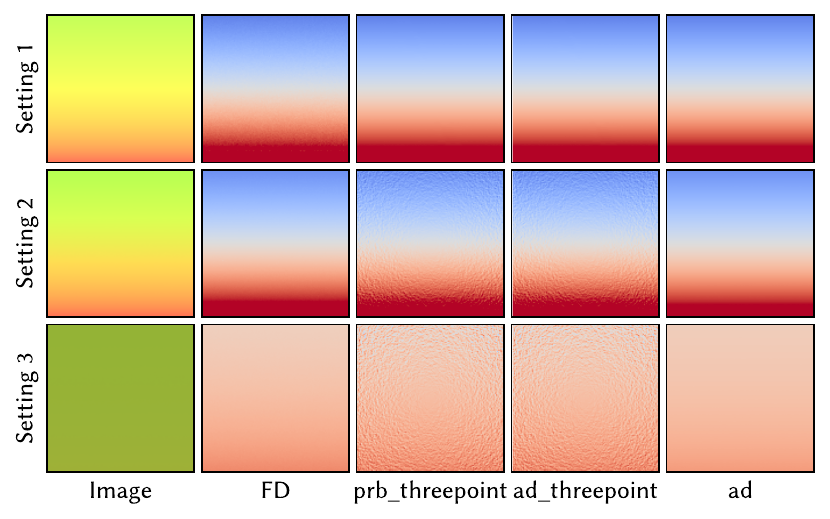}
    \caption{Three scenes with a backward-moving rectangle -- spatially-varying emission (first row), spatially varying BSDF (second row), and indirect illumination (third row) -- are rendered (first column). The derivatives of the red channel are displayed: finite differences (FD) (second column), PRB using the three-point form (third column), automatic differentiation (AD) using the three-point form (fourth column), and automatic differentiation (AD) using the spherical form (last column). The derivatives of the three-point form integrators are close to the FD reference, but they differ from the spherical form because: (1) the estimates have different noise patterns, and (2) the scenes have discontinuities in the three-point form, which leads to a small bias.
    }
    \label{fig:case_1_threepointform}
    \iftoggle{preprint}{
        \Description[]{} %
    }{}
\end{figure}

\subsection{Case Study -- Revisited} 

We first revisit the case study from \cref{sec:case_1} and test our integrators in the three settings (\cref{fig:case_1_threepointform}). In contrast to existing implementations (recall \cref{fig:case_1}), the integrators based on the three-point form produce non-zero derivatives in the last setting, which not only agree with the finite difference reference, but PRB and AD match exactly, consistent with the theory. Both, direction and point sampling produce similar results that differ, for example, by the noise pattern. One reason is that individual samples contribute differently to the derivative in both forms, so equal sample counts are not necessarily comparable (see \cref{sec:detached_positions}, Paragraph~\hyperref[par:contribution_of_samples]{Contribution of individual samples}).

The unbiased solution (\texttt{prb\_threepoint}) has an overhead over the biased solution (\texttt{prb}), which, however, is expected because the AD graph within an iteration is more complex as the previous and next surface intersections are now considered.

\subsection{Unique Discontinuities} 

\begin{figure}
    \centering
    \includegraphics[width=\linewidth]{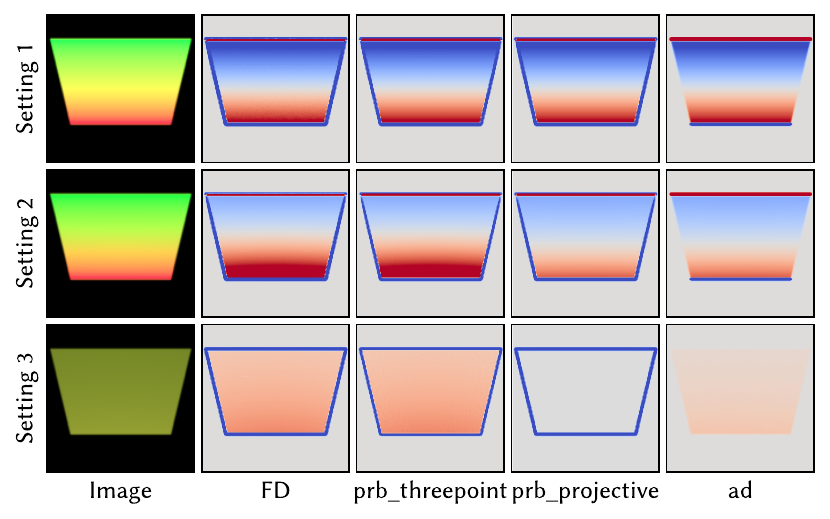}
    \caption{Three scenes with a backward-moving scaled rectangle -- spatially-varying emission (first row), spatially varying BSDF (second row), and indirect illumination (third row) -- are rendered (first column). The derivatives of the red channel are displayed: finite differences (FD) (second column), PRB integrator using three-point formulation (third column), the projective sampling variant of PRB (fourth column), and automatic differentiation (AD) using spherical formulation (last column). Only the integrator based on the three-point form computes the correct derivatives.}
    \label{fig:case_1_full_visibility}
    \iftoggle{preprint}{
        \Description[]{} %
    }{}
\end{figure}

We have so far assumed that the integrand is continuous, but moving geometry can certainly lead to discontinuous integrands. What constitutes a ``discontinuity'' depends solely on the integration domain: when integrating over directions, the radiance collected along a sampled direction can change discontinuously as an object moves into or out of the ray. When integrating over surface points, the collected radiance can change discontinuously as the \emph{visibility} of a surface point changes. The two perspectives are not generally interchangeable and a scene with no discontinuities in one form may have discontinuities in the other. For example, we initially suggested that the first case study scene with a moving slab covering the camera's field of view was free of discontinuities (\cref{sec:case_1}) but this is only true for the spherical integral. In fact, all scenes from \cref{sec:case_1} have discontinuities in the three-point form, which leads to bias in the derivatives in \cref{fig:case_1_threepointform}.

We have modified all scenes to be free of discontinuities in the three-point form, which is achieved by reducing the size of the slab and the emitter, making the slab fully visible to the camera and the emitter fully visible to each point on the slab. In these settings, the three-point form yields unbiased derivatives (\cref{fig:case_1_full_visibility}). However, Settings 2 and 3 now have discontinuities when sampling the (hemi)sphere, which noticeably affects the result of the AD integrator (\verb|ad|) based on detached sampling of directions (\cref{sec:detached_directions}). The projective sampling integrator is susceptible to the same discontinuities (see Setting 2).

\subsection{Radiative Backpropagation for Interior Path Derivatives}

\begin{figure}
    \centering
    \includegraphics[width=\linewidth]{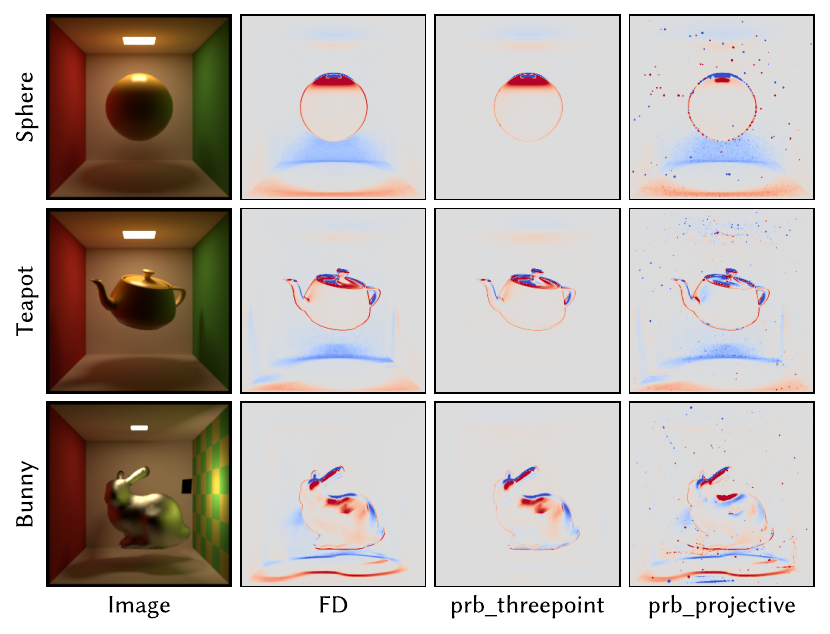}
    \caption{Implementations of path-space differentiable rendering combine interior derivatives computed with radiative backpropagation (RB) and boundary derivatives (\texttt{prb\_projective}). Their interior derivatives do not match the finite difference reference (FD). An RB implementation of the surface integral (\cref{sec:detached_positions}) produces the correct interior derivatives (\texttt{prb\_threepoint}).}
    \label{fig:case_2_forward}
    \iftoggle{preprint}{
        \Description[]{} %
    }{}
\end{figure}

Although we have entirely disregarded discontinuities in the derivations so far, it is straightforward to combine our approach with discontinuity handling using the framework of path-space differentiable rendering: Zhang et al.~\shortcite{Zhang:2020:PSDR} show that the differential path integral, i.e., a differential integral over the product spaces of surface integrals, can be decomposed into an \emph{interior} and a \emph{boundary} term. Recent work has been mainly focused on the boundary component because it requires sampling visibility discontinuities~\cite{Zhang:2020:PSDR, Zhang:2023:Projective}, whereas the interior component has received less attention as it is simply ``differentiating under the integral'' and generally assumed to be straightforward to compute. This is true when using automatic differentiation, but, computing the interior term with radiative backpropagation requires the general theory for non-static geometry: our approach for detached path replay backpropagation in the three-point form (\cref{sec:detached_positions}) computes an unbiased estimate of the interior term in RB-style. In other words, combining the approach from \cref{sec:detached_positions} with discontinuity handling is as simple as choosing a suitable method for computing the boundary term and adding the result to our estimate of the interior term. Existing implementations that combine the boundary term with an RB-style interior term compute an \emph{incorrect} interior term, based on direction sampling, which manifests as severe bias, e.g. with inverted derivative signs (\cref{fig:case_2_forward}).

\begin{figure}
    \centering
    \includegraphics[width=\linewidth]{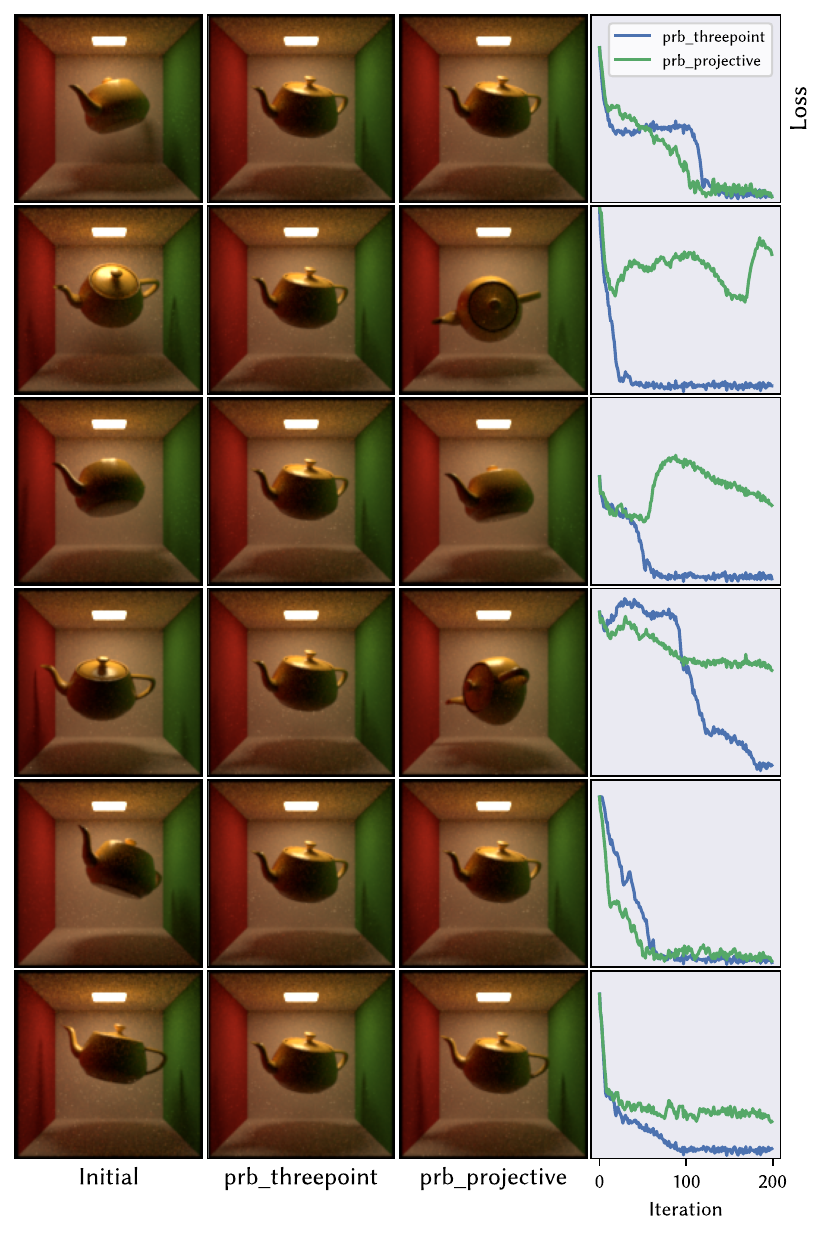}
    \caption{A simple inverse rendering task -- recovering the 6D pose (position and orientation) of the teapot from a single image -- fails if the interior derivatives are incorrect, even if discontinuities are handled (\texttt{prb\_projective}). An RB-based implementation of the surface integral (\cref{sec:detached_positions}) converges each time, despite not explicitly handling discontinuities (\texttt{prb\_threepoint}).}
    \label{fig:case_2_pose}
    \iftoggle{preprint}{
        \Description[]{} %
    }{}
\end{figure}

\paragraph*{A simple application}

Despite the presence of discontinuities, the interior component is central for convergence, even in the simplest inverse rendering applications: we test implementations on a monocular pose estimation task, where the orientation and position of an object are to be inferred from a single image. The objective is to minimize the mean absolute error between a target image and the rendering. Current implementations combining an RB-style interior term with the boundary term (here estimated with projective sampling~\cite{Zhang:2023:Projective}) often fail to converge in the given time whereas the three-point integrator convergences successfully, even without explicit discontinuity handling (\cref{fig:case_2_pose}).

\paragraph*{Combining boundary and RB interior derivatives}

\begin{figure*}
    \centering
    \includegraphics[width=0.495\linewidth]{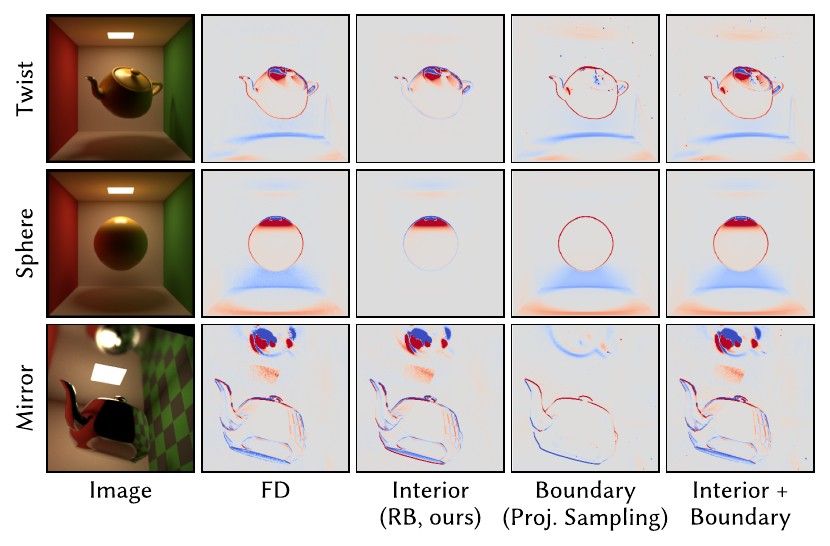}
    \includegraphics[width=0.495\linewidth]{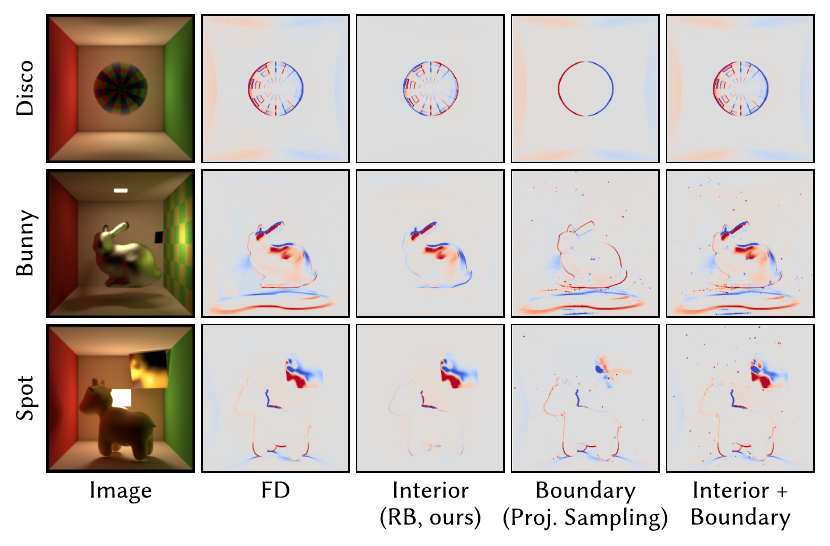}
    \caption{For path-space differentiable rendering, the interior component of the derivative can be computed with radiative backpropagation (RB) in the surface integral form (\cref{sec:detached_positions}), and the boundary component with projective sampling~\cite{Zhang:2023:Projective}. We combine both in a proof-of-concept implementation (``Interior + Boundary''). While this implementation matches the finite difference (FD) reference more closely than existing ones (compare \cref{fig:case_2_forward}), the boundary component still leads to bias in the result (e.g. bottom of \textsc{Bunny}). The shown derivatives are w.r.t.\ a translation of the center object.}
    \label{fig:case_2_projective_fix}
    \iftoggle{preprint}{
        \Description[]{} %
    }{}
\end{figure*}

As a proof-of-concept, we have implemented an integrator that combines projective sampling~\cite{Zhang:2023:Projective} with our RB interior derivatives (\cref{fig:case_2_projective_fix}). While the combination of derivatives matches the finite difference reference more faithfully, there is still bias in the result, although this combination should be unbiased. The decomposition of the interior and boundary contributions in \cref{fig:case_2_projective_fix} reveals that a systematic bias originates from the boundary component estimated with the official implementation of projective sampling in \mitsuba (see bottom of \textsc{Bunny}). It is unclear, whether this is a conceptual issue or related to the implementation. However, since the boundary term is unrelated to the findings in this paper and beyond its scope, we defer further investigation to future work.

\section{Discussion}
\label{sec:discussion}

The original theory underlying radiative backpropagation assumes static geometry and spherical integrals. We have complemented the theory by deriving the differential radiance equations for non-static geometry for different parameterizations of the rendering integral.

A key observation is that differential radiance is not only transported like normal radiance but differential radiance is also emitted from moving geometry. Depending on the parameterization of the rendering integral, this emission term becomes non-local, so where previously derivatives could simply be computed from local path geometry, the motion of the entire light path must to be taken into account. We have shown that the derivatives become local when the rendering integral is reparameterized over surfaces, leading to practical implementations for non-static geometry.

\subsection{Related Work}

A combination of (detached) radiative backpropagation and non-static geometry with discontinuity handling~\cite{Loubet2019, Bangaru2020} has previously been attempted by Zeltner et al.\shortcite{Zeltner2021}. They transform the differential scattering equation \eqref{eq:diff_scattering_nimier} but do not consider the implications for the differential transport equation \eqref{eq:diff_transport_nimier}, therefore arriving at an approach that is still local. Amending their derivation with our findings should lead to a correct combination of the techniques for spherical integrals. But, as shown, the result will be significantly more complex due to the non-local terms.

The theory of radiative backpropagation by Stam~\shortcite{Stam:2020:Adjoint} also builds on the three-point form, as Section~\ref{sec:detached_positions}, albeit in a different (theoretical) framework, and it does not consider movement of geometry with scene parameters.

The surface parameterization in \cref{sec:detached_positions} corresponds to the material-form parameterization introduced for path-space differentiable rendering by Zhang et al.~\cite{Zhang:2020:PSDR}. They rely on automatic differentiation for the interior term, whereas we show how these derivatives can be computed using radiative backpropagation.

In concurrent work, Finnendahl and Worchel et al.~\cite{Finnendahl_and_Worchel:2025:DiffAcousticPT} have extended path replay backpropagation to differentiable time-resolved rendering. Since path geometry affects the transport time of energy, the time-domain derivatives must account for perturbations caused by non-static surfaces. This is accomplished using PRB in the three-point form, as presented in \cref{sec:detached_positions}.

\subsection{Impact on Implementations}

We have shown that the \verb|prb| and \verb|prb_projective| integrators in \mitsuba (version 3.6.4) are affected by bias if geometry is non-static. The now deprecated \verb|prb_reparam| integrator is very likely affected as well because it builds on the findings by Zeltner et al.~\cite{Zeltner2021}, who arrive at a local formulation of derivatives that does not correctly account for the differential transport. Results relying on the affected integrators (e.g. for academic publications) should very likely be re-validated.

Physically-based renderers that do not rely on radiative backpropagation, such as \redner~\cite{redner} (based on work by Li et al.~\cite{Li2018}) or \psdr~\cite{psdr}, are unaffected by this bias. While our RB implementation is based on \mitsuba, we believe it also provides a useful reference point for other frameworks.

\section{Acknowledgements}
We would like to thank%
\iftoggle{review}{
    anonymous researcher for reviewing this \iftoggle{preprint}{report}{paper}.
}{
    Wenzel Jakob for providing feedback on this paper and the historical context for the (P)RB implementations, and we thank the anonymous reviewers for their helpful suggestions. This work was funded by the European Research Council (ERC) under the European Union’s Horizon 2020 research and innovation programme (Grant agreement No. 101055448, ERC Advanced Grand EMERGE).
}

\bibliography{bibliography}

\begin{appendix}

\section{An Attached Approach for Detached Sampling}
\label{sec:attached_as_detached}

Vicini et al.~\shortcite{Vicini2021} consider the recursion of $\dfdx{L_o}{\posS}$ in Eq.~\eqref{eq:diff_transport_position} for the attached case of PRB but with different reasoning. The extended version of \cref{eq:attached}, without the emission $L_e$, reads \cite[Supplementary Material]{Vicini2021}:
\begin{alignat}{2}
\begin{split}
   \dfdx{L_o}{\pi}(\posS,\solid) = \int\limits_{\mathcal{U}^2} &L_i\big(\posS,T(\mv{u},\pi)\big) \,\dfdx{\left[ \frac{f\big(T(\mv{u}, \pi), \pi \big)}{p\big(T(\mv{u}, \pi), \pi\big)} \right]}{\pi}\\
   &+ \dfdx{\left[L_i\big(\posS,T(\mv{u},\pi_0)\big)\right]}{\pi}\, \frac{f\big(T(\mv{u}, \pi), \pi \big)}{p\big(T(\mv{u}, \pi), \pi\big)}\\
   &+ \dfdx{L_i}{\solid}\big(\posS,T(\mv{u},\pi_0)\big)\, \dfdx{\big[T(\mv{u},\pi)\big]}{\pi}\, \frac{f\big(T(\mv{u}, \pi), \pi \big)}{p\big(T(\mv{u}, \pi), \pi\big)} \, \text{d} \mv{u}.
\end{split}
\end{alignat}
The reparameterization $T$ introduces a new term that involves $\dfdx{L_i}{\solid}(\posS,\solid)$, which can be expanded using the transport equation
\begin{multline}
   \dfdx{L_i}{\solid}(\posS,\solid) = \dfdx{L_o}{\solid}\big(\hitS(\posS,\solid_0),\solid\big) + \\
   \dfdx{L_o}{\posS}\big(\hitS(\posS_0,\solid_0),\solid_0\big)\dfdx{\hitS}{\solid}(\posS,\solid),
\end{multline}
where $\solid_0 = \solid$ and $\posS_0 = \posS$. For attached sampling, one must therefore compute $\dfdx{L_o}{\posS}$ from \cref{eq:diff_scattering_position}, not because geometry moves but because the sample directions depend on $\pi$ and so do the intersection points.%

Vicini et al.~\shortcite{Vicini2021} compute $\dfdx{L_o}{\posS}$ using forward-mode differentiation, and, by our previous arguments, this is also required in the detached case if geometry is non-static. In their Jacobian framework, the Monte Carlo integration can be written as
\begin{alignat}{2}
   &h^{(N)}(\pi,L_0,\beta_0,\solid_0,\posS_0), \text{with}\\
   &h(\pi,L,\beta,\solid,\posS) = \big[\pi, L+\beta L_e, \beta f, \solid_i, \hitS(\posS,\solid)\big],
\end{alignat}
where $N$ is the maximum path depth and $(L_0=0,\beta_0=1,\solid_0,\posS_0)$ is the start configuration. The PDF $p$ is omitted for clarity and because it does not participate in differentiation in the detached case. For the derivative of each intersection (i.e., invocation of $h$) one has the reduced Jacobian (reduced as it ignores $\pi$)
\begin{equation}
   J_h = \begin{bmatrix}
        1 & L_e & \beta\dfdx{L_e}{\solid} & \beta\dfdx{L_e}{\posS}  \\
        0 &   f &   \beta\dfdx{f}{\solid} &   \beta\dfdx{f}{\posS}  \\
        0 &   0 &                       0 &                     0  \\
        0 &   0 &                       0 &     \dfdx{\hitS}{\posS}  \\
   \end{bmatrix}.
\end{equation}
Given that Vicini et al.~\shortcite{Vicini2021} parametrize $(\posS,\solid)$ as a single ray, this resembles the code presented in their Listing 3. To convert the pseudocode to a detached estimator for non-static geometry, one only needs to detach the pdf and $\solid_i$ ($=\solid'$ in Listing 3).
Unfortunately, there is no publicly available implementation, that could be used for validation, and we did not re-implement the approach.

\section{Derivation of the Three-Point Form Pseudocode}
\label{sec:appendix}

To verify that all terms of the derivative are computed, we take the explicit three-point form
\begin{equation}
    \begin{split} 
    L_o&(\p, \solid, \pi) = L_e(\p, \solid, \pi) \\ 
    &\begin{alignedat}{4}
&& +\int_\mathcal{M} L_i\big(&\p, \solidn(\p,\pn,\pi), \pi\big) \\ 
    && f\big(&\p, \solid, \solidn(\p,\pn,\pi), \pi\big)\, D(&\p, \pn, \pi)\,\text{d}\pn,
    \end{alignedat}
    \end{split}
\end{equation}
with $\solidn(\p,\pn,\pi)=\Sp(\p,\pi) \rightarrow \Sp(\pn,\pi)$. As we now sample surface points $\pn$ and construct the outgoing direction $\solid$, the transport equation simplifies to
\begin{equation}
    \begin{split}
    L_i\big(\p&, \solidn(\p,\pn,\pi), \pi\big) = \\
    &\begin{alignedat}{4} 
    &&L_o&\left(\getpn\left(\p, \solidn(\p,\pn,\pi), \pi\right), \solidn(\pn,\p,\pi), \pi\right) \\ 
    &&= L_o&\left(\pn, \solidn(\pn,\p,\pi), \pi\right),
    \end{alignedat}
    \end{split}
\end{equation}
as $\getpn\big(\p, \solidn(\p,\pn,\pi), \pi\big) = \pn$. Thus, the reparameterization eliminates the dependency on $\pi$ for $\p$, but introduces a new dependency on $\pi$ for the direction $\solid$ (compare with \cref{eq:transport}).
The derivative with respect to $\pi$ is therefore given by
\begin{equation}\label{eq:three-point-derivative}
\begin{split}
    \dfdx{&L_o}{\pi}(\p, \solid,\pi) = \dfdx{L_e}{\pi}(\p, \solid,\pi)\\
    &
    \!\begin{alignedat}{6}
        + \int_{\mathcal{M}} \dfdx{L_o}{\pi}\left(\pn, \solidn(\pn,\p,\pi),\pi\right)& \, &&f \, &&D(\p, \pn, \pi) \\
        +\, L_o\left(\pn, \solidn(\pn,\p,\pi),\pi\right)& \, \dfdx{&&f}{\pi} \, &&D(\p, \pn, \pi) \\
        +\, L_o\left(\pn, \solidn(\pn,\p,\pi),\pi\right)& \, &&f \, \dfdx{&&D}{\pi}(\p, \pn, \pi) \\
        \text{d}\pn,
    \end{alignedat}\\
\end{split}
\end{equation}
where $f = f\left(\p, \solid, \solidn(\p,\pn,\pi),\pi\right)$. The consecutive derivative of $L_o$ reveals the dependency on $\pi$ for the direction $\solid$:
\begin{equation}\label{eq:three-point-recursion-derivative}
\begin{split}
    \dfdx{&L_o}{\pi}\left(\pn, \solidn(\pn,\p,\pi), \pi\right) = \dfdx{L_o}{\pi}\left(\pn, \solidn(\pn,\p,\pi_0), \pi\right)\\
    &+\dfdx{L_o}{\omega}\left(\pn, \solidn(\pn,\p,\pi), \pi\right)\,\dfdx{}{\pi}\left[\solidn(\pn,\p,\pi)\right].
\end{split}
\end{equation}
We therefore need the derivative of $L_o$ with respect to $\solid$:
\begin{equation}
\begin{split}
    &\dfdx{L_o}{\solid}(\p, \solid, \pi) = \dfdx{L_e}{\solid}(\p, \solid,\pi)\\
    &
    \!\begin{alignedat}{6}
        + \int_{\mathcal{M}} L_o\left(\pn, \solidn(\pn,\p,\pi),\pi\right)& \, \dfdx{&&f}{\solid}\left(\p, \solid, \solidn(\p,\pn,\pi),\pi\right) \, \\
        D(\p, \pn, \pi)\,\text{d}\pn.
    \end{alignedat}
\end{split}
\end{equation}
This reveals the locality enabled by the reparameterization as mentioned in \cref{sec:breakingtherecursion}: the derivative $\dfdx{L_o}{\solid}$ is not recursive in contrast to $\dfdx{L_o}{\p}$.

In \verb|sample_path_adjoint| in \cref{lst:pseudocode}, we keep track of the three surface points that affect the path geometry around the current intersection point (i.e., previous $\mv{p}_p$, current $\mv{p}_c$, and next $\mv{p}_n$) and accumulate the contribution of a single position sample over three loop iterations: within a single iteration of the for loop, we construct the direction $\solid$, attached to the AD graph. In this way, we calculate $\dfdx{L_e}{\pi}(\p, \solid)$ and $\dfdx{L_e}{\solid}(\p, \solid)\dfdx{}{\pi}[\solidn(\pn,\p,\pi)]$ in line 10. The same is true for the two derivative terms related to the BSDF ($\dfdx{f}{\pi}$ and $\dfdx{f}{\solid}$) in line 24f. The term containing the derivative of the determinant is calculated in line 14. In the next iteration, the recursive term $\dfdx{L_o}{\pi}$ is evaluated with the updated $\solid$. Therefore, all terms of \cref{eq:three-point-recursion-derivative} are computed by \verb|sample_path_adjoint|.

\end{appendix}

\end{document}